\documentclass{article}

\usepackage{arxiv}

\usepackage[utf8]{inputenc} 
\usepackage[T1]{fontenc}    
\usepackage{hyperref}       
\usepackage{url}            
\usepackage{booktabs}       
\usepackage{amsfonts}       
\usepackage{nicefrac}       
\usepackage{microtype}      
\usepackage{lipsum}		
\usepackage{graphicx}
\usepackage{natbib}
\usepackage{doi}

\usepackage{amsthm,amsmath}

\RequirePackage{hyperref}

\usepackage{calc}

\usepackage{caption}

\PassOptionsToPackage{table}{xcolor}
\usepackage[table]{xcolor}
\usepackage{multicol}
\usepackage{colortbl,array}
\usepackage{bigstrut}
\usepackage{multirow}
\usepackage{makecell}
\usepackage{subfig}
\usepackage{arydshln}
\usepackage{units}
\usepackage{booktabs} 
\usepackage{tablefootnote} 

\usepackage{float}
\usepackage{amssymb}
\usepackage{pifont}
\usepackage{ragged2e}

\newcommand{\xmark}{\ding{53}}%

\usepackage{tikz}

\newcommand\submittedtext{%
  \footnotesize This work has been submitted to the IEEE for possible publication. Copyright may be transferred without notice, after which this version may no longer be accessible.}

\newcommand\submittednotice{%
\begin{tikzpicture}[remember picture,overlay]
\node[anchor=south,yshift=10pt] at (current page.south) {\fbox{\parbox{\dimexpr0.65\textwidth-\fboxsep-\fboxrule\relax}{\submittedtext}}};
\end{tikzpicture}%
}

\title{w2v-SELD: A Sound Event Localization and Detection Framework for Self-Supervised Spatial Audio Pre-Training}

\author{ \href{https://orcid.org/0000-0002-3942-6418}{\includegraphics[scale=0.06]{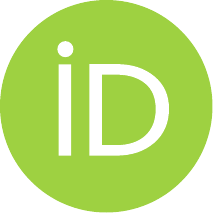}\hspace{1mm}Orlem Lima dos Santos} \\
	Department of Computer Engineering \\ and Industrial 
        Automation\\
	University of Campinas\\
	Sao Paulo, Campinas \\
	\texttt{o211501@dac.unicamp.br} \\
	\And
	\href{https://orcid.org/0000-0002-8118-4213}{\includegraphics[scale=0.06]{orcid.pdf}\hspace{1mm}Karen Rosero} \\
	Department of Electrical and Computer Engineering\\
	The University of Texas at Dallas\\
	Texas, Richardson \\
	\texttt{kgr220000@utdallas.edu} \\
        \And 
        \href{https://orcid.org/0000-0002-5652-0852}{\includegraphics[scale=0.06]{orcid.pdf}\hspace{1mm}Roberto de Alencar Lotufo} \\
	Department of Computer Engineering \\ and Industrial 
        Automation\\
	University of Campinas\\
	Sao Paulo, Campinas \\
	\texttt{lotufo@unicamp.br} \\
}

\renewcommand{\shorttitle}{ }

\hypersetup{
pdftitle={w2v-SELD: A Sound Event Localization and Detection (SELD) Framework for Self-Supervised Spatial Audio Pre-Training},
pdfsubject={eess.AS},
pdfauthor={Orlem Lima dos Santos, Karen Rosero},
pdfkeywords={Sound Event Localization and Detection, Self-Supervised Learning, wav2vec 2.0, Spatial Audio},
}

\begin{document}
\maketitle

\submittednotice

\begin{abstract}
\justifying
Sound Event Detection and Localization (SELD) constitutes a complex task that depends on extensive multichannel audio recordings with annotated sound events and their respective locations. In this paper, we introduce a self-supervised approach for SELD adapted from the pre-training methodology of wav2vec 2.0, which learns representations directly from raw audio data, eliminating the need for supervision. By applying this approach to SELD, we can leverage a substantial amount of unlabeled 3D audio data to learn robust representations of sound events and their locations. Our method comprises two primary stages: pre-training and fine-tuning. In the pre-training phase, unlabeled 3D audio datasets are utilized to train our w2v-SELD model, capturing intricate high-level features and contextual information inherent in audio signals. Subsequently, in the fine-tuning stage, a smaller dataset with labeled SELD data fine-tunes the pre-trained model. Experimental results on benchmark datasets demonstrate the effectiveness of the proposed self-supervised approach for SELD. The model surpasses baseline systems provided with the datasets and achieves competitive performance comparable to state-of-the-art supervised methods. The code and pre-trained parameters of our w2v-SELD model are available in \href{https://github.com/Orlllem/seld_wav2vec2.git}{this repository}. 
\end{abstract}

\keywords{Sound Event Localization and Detection \and Self-Supervised Learning \and wav2vec 2.0 \and Spatial Audio}

\section{Introduction}


The Sound Event Detection and Localization (SELD) task is a cutting-edge application to analyze complex acoustic scenes. Inspired by the human auditory system, which processes sound coming from all our surroundings, the SELD task relies on spatial audio which preserves or recreates the three-dimensional perception of natural sound using multichannel Ambisonic recordings. From a computational perspective, SELD encompasses two tasks: sound event detection (SED) and direction of arrival (DOA) estimation, which is also referred to as sound source localization (SSL)~\cite{guirguis2021seld}. The SED task involves identifying the start and end times of every sound event and the corresponding class, while the DOA estimation predicts the respective three-dimensional coordinates from which the sounds are coming. Therefore, the SELD task combines three essential components of sound analysis: temporal identification, spatial localization, and semantic labeling of sound events throughout time. The development of computational solutions to perform SELD in real-time opens up new avenues for innovation and automation in various fields where the accurate analysis of sounds is of utmost importance, such as environmental monitoring, security, entertainment, smart meeting rooms, and transportation~\cite{adavanne2018sound, guizzo2021l3das21}.  




While SED has generally been treated as a multi-label classification problem, the DOA estimation has evolved from the use of parametric-based approaches into Deep Neural Network (DNN)-based approaches. This tendency responds to the limitations of parametric-based methods in capturing complex and non-linear relationships in the data. In contrast, DNN-based methods have demonstrated the ability to accurately estimate the DOA, even in highly complex and noisy scenarios. The SELDnet~\cite{adavanne2018sound} was the first DNN model that gained relevance in the field because it performs the SELD task by training a single  Convolutional Recurrent Neural Network (CRNN) with two branches: one for SED and the other for DOA estimation. Further works in the area include the use of Temporal Convolutional Networks (TCN)~\cite{guirguis2021seld}, ensemble models~\cite{kapka2019sound}, and tailored data augmentation techniques~\cite{park2019specaugment}. 


Despite the good performance of state-of-the-art models for SELD, the limited amount of multichannel recordings contained in spatial audio datasets hinders the models from reaching better metrics. The lack of extensive datasets for the SELD task can be explained by the need for special equipment to record spatial audio, such as microphone arrays, but also because of the arduous labeling process, in which the timestamps, location, and class of every sound event should be accurately annotated to be used on supervised learning approaches.   

In contrast, the field of speech recognition has experienced significant advancement with the introduction of Transformer-based models coupled with a self-supervised pre-training framework, especially the wav2vec 2.0 model~\cite{baevski2020wav2vec}. This enables the model to learn general representations from vast amounts of unlabeled audio data during pre-training. The pre-training step is crucial as it allows the model to establish a strong foundation of knowledge before being fine-tuned on transcribed speech (labeled data).   

We have identified a lack of usage of pre-trained models for SELD, that even though were not specifically trained for this task, can leverage vast amounts of information from unlabeled audio data from datasets such as Librispeech (LS-960) and LibriVox (LV-60k). In response to this shortfall, we have adapted the architecture of wav2vec 2.0 to accommodate a multichannel input and an output containing the SED and DOA predictions. We refer to this model as w2v-SELD. Additionally, we explore the gap between using the model being pre-trained on mono-channel speech data compared with using unlabeled spatial audio. Subsequently, the pre-trained model is fine-tuned into a specific domain using transfer learning. This approach also avoids training several models independently for a specific task before ensembling them for improved predictions.




Our results demonstrate that modifying w2v-SELD to perform SED and DOA predictions over an entire frame results in a better performance than using the original sequential approach of the model. We also verify a $20\%$ improvement of the SELD\textsubscript{score} metric when using pre-trained weights instead of training the model from scratch. Furthermore, pre-training w2v-SELD on unlabeled spatial audio improves the SELD\textsubscript{score} metric by $40\%$ compared with pre-training on mono-channel audio. Lastly, we compare the performance of our w2v-SELD approach with the baseline systems provided for each dataset, and with the state-of-the-art systems. The SELD\textsubscript{score} improved by $66\%$ with respect to the baseline system performance and reached close state-of-the-art performance using raw spatial audio as input instead of relying on spectrograms, phase, or intensity vectors.


The main contributions of this work are as follows:
\begin{itemize}

    \item Development of a Self-Supervised Learning (SSL) approach for SELD, based on the wav2vec 2.0 pre-training framework;
    \item Avoid labeling extensive amounts of spatial audio datasets required to train supervised SELD models, thereby enhancing the practicality and affordability of the proposed method;
    \item Adaptation of the fine-tuning process to encompass SED and DOA estimation to predict on the frame-level;
    \item Thorough examination and evaluation of the effectiveness of our w2v-SELD model pre-trained on different datasets.
\end{itemize}

We also share the code\footnote{The code is available at \url{https://github.com/Orlllem/seld_wav2vec2.git}} built upon fairseq \cite{ott2019fairseq} and the pre-trained weights BASE and LARGE of w2v-SELD to be used in SELD tasks of limited size.

The paper is structured as follows: In Section \ref{sec:related}, we provide a comprehensive literature review concerning the domains of SELD for spatial audio. In Section \ref{sec:background}, we expound upon the theoretical foundations of SSL. Section \ref{sec:method} details the methodology employed in our research, while Section \ref{sec:setup} describes the experimental setup. Section \ref{sec:results} presents the outcomes of our experimental endeavors. Finally, critical points are discussed, and conclusive remarks are addressed in Sections \ref{sec:discussion} and \ref{sec:conclusion}, respectively.

\section{Related Works}\label{sec:related}

In this section, we provide a summary of key contributions and advancements in the field of SELD as reported in the literature, highlighting the major trends and challenges encountered in this area of research.

\begin{figure}[!htb]
    \centering
    \includegraphics[width=0.55\linewidth]{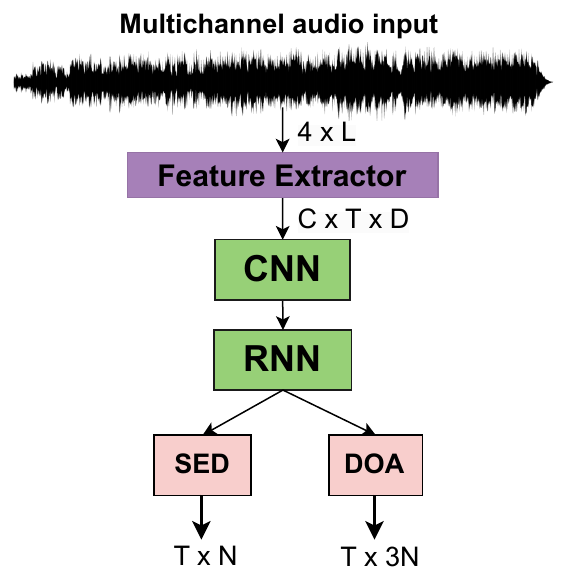}
    \caption{\label{fig:seldnet}Illustration of the SELDnet model. Adapted from \cite{adavanne2018sound}.}
\end{figure}

The SELDnet, proposed by \cite{adavanne2018sound}, was the first effective Neural Network (NN) to accomplish both SED and DOA estimation tasks without using parametric methods, then, it is often considered as a baseline for the SELD task. The model jointly deals with the SELD problem through a shared CRNN followed by two branches: one for SED and one for DOA estimation, as shown in Figure \ref{fig:seldnet}. The phase and magnitude spectrograms are computed for each channel of the spatial audio input, resulting in a sequence of spectrogram frames that are fed into the model as input features. Later, the CNN blocks reduce the dimensionality and extract  relevant characteristics of the sequence, while the RNN blocks capture temporal correlations. Next, the SED branch performs a multi-label classification task, which enables the network to simultaneously estimate the presence of up to three active sound events for each frame. On the other hand, the DOA branch performs a multi-output regression task, for which each sound event class is associated with three regressors that estimate the 3D Cartesian coordinates $(x, y, z)$ of the DOA on a unit sphere surrounding the microphone \cite{adavanne2018sound}. Since SELDnet can detect multiple concurrent sound events and their locations, it is a polyphonic system that maps the temporal activity and DOA trajectories for each sound event class. Inspired by SELDnet, \cite{guirguis2021seld} replaced the Recurrent Neural Network (RNN) layers in the SELDnet with a TCN to improve the performance while reducing the computational cost of the model. In a similar vein, \cite{g-seld} combined TCN and RNNs to improve the NN model, and used a Gammatone filterbank to reduce the frequency bins of the spectrogram in the preprocessing stage.

The SELD task has been considered in several releases of the DCASE Challenge, which has introduced different complexities such as reverberation and interference sounds. State-of-the-art approaches for DCASE 2019 and 2020 for SELD are based on strong data augmentations and ensemble models. For DCASE 2019, authors of \cite{kapka2019sound} implemented the SpecAugment \cite{park2019specaugment} technique for data augmentation and trained four SELDnets sequentially to achieve the best results of the challenge. For DCASE 2020, \cite{Du2020_task3_report} introduced an ensemble of DNNs, including ResNet \cite{he2016} and Xception \cite{chollet2017}, both combined with (Gated Recurrent Units) GRUs or factorized Time Delay Neural Networks (TDNNF). These ensemble models are able to extract high-level feature representations and temporal context representations. Further than using SpecAugment for data augmentation, the authors generated new Ambisonics audio samples by mixing the non-overlapping samples in the development dataset and transforming them into the spatial audio format. 

In the subsequent DCASE 2021 challenge, \cite{Shimada_SONY_task3_report} proposed an ensemble method known as \textit{Activity-Coupled Cartesian Direction of Arrival} (ACCDOA), which emerged as a prominent baseline approach in the SELD field. ACCDOA eliminated the SED branch, which has traditionally been used in SELD methods, make the SED a function of the length of the Cartesian DOA vector \cite{shimada2021accdoa}. The introduction of ACCDOA has significantly influenced the SELD field by demonstrating the potential for novel and innovative approaches to address the problem with a single branch. 


In parallel with the advancements introduced by the use of attention mechanisms in the area of speech processing, \cite{schymura2021} proposed an attention-based deep recurrent neural network for localizing sound events in spatial audio, named ADRENALINE. The attention mechanism in ADRENALINE captures long-term dependencies on the feature representation when localizing sound events. ADRENALINE surpassed the DOA estimation obtained with SELDnet. Further advancements in the field of DOA estimation include the probabilistic localization of sounds with transformers, named PILOT \cite{schymura2021b}, in which the recurrent structures were replaced with a transformer-based model able to handle temporal dependencies in sequence data. Both approaches omitted the SED task in their frameworks.  These techniques rely solely on annotated datasets, necessitating significant manual effort for the annotation process. This reliance on labeled data constrains the models' capacity for generalization and performance, restricting their applicability to only those datasets with available labels.

The field of speech recognition has experienced a significant advancement with the introduction of self-supervised pre-training frameworks, which enable the model to leverage vast amounts of unlabeled audio data. \cite{scheibler2022sound} adapted a Self-Supervised Audio Spectrogram Transformer (SSAST) for the SELD task. They used the SSAST weights that were pre-trained on mono-channel utterances from the Audioset dataset \cite{gemmeke2017audio}, which comprises $4,971$ hours of audio stored in 10-second segments. Then, the model was fine-tuned on the STARSS22 dataset \cite{politis2022starss22}, which contains 4.9 hours of spatial audio specific to the SELD task. Even though SSAST for SELD introduced the benefit of using pre-trained networks and fine-tuning them for the SELD task, the option of pre-training the model with unlabeled spatial audio remains unexplored.


 In contrast, the SSL approach proposed for our w2v-SELD model adapts the wav2vec2.0 pre-training framework to learn from spatial audio. Moreover, we offer a comparative analysis between using mono-channel or multichannel audio for pre-training.  

\section{Theoretical Background}\label{sec:background}


This section explores the key concepts, techniques, and motivations behind self-supervised Transformers for audio and speech representations. Specifically, we describe the key aspects of the wav2vec 2.0 model and its pre-training methodology, which are adapted for the SELD task in this study.



Self-supervised Transformers are able to learn accurate audio and speech representations by training on large amounts of unlabeled audio data. The learned audio representations have been used in downstream tasks such as speaker verification \cite{chen2022}, speech recognition \cite{baevski2020wav2vec}, or emotion recognition \cite{chen2023}. Specifically, this approach has achieved state-of-the-art performance in speech recognition, demonstrating remarkable reductions in the word error rates.  

\subsection{wav2vec 2.0 model}

The wav2vec 2.0 model, proposed by \cite{baevski2020wav2vec}, is one of the current state-of-the-art frameworks for unsupervised speech representations. Through training on a massive amount of audio data, the model learns to extract high-level features, such as phonemes and sub-word units, directly from raw waveform data. 


The model comprises three main components: 1) feature encoder, 2) context network, and 3) quantization module, which will be briefly described.



\begin{figure}[!htb]
    \centering
    \includegraphics[width=0.65\linewidth]{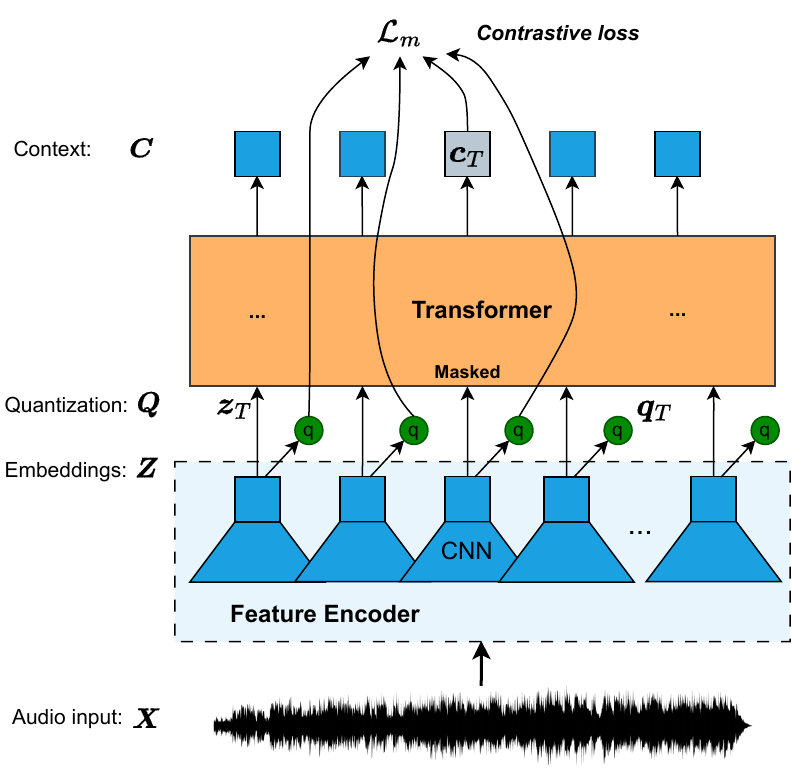}
    \caption{Illustration of the wav2vec 2.0 model. Adapted from \cite{baevski2020wav2vec}.}
    \label{fig:wav2vec2}
\end{figure}

The feature encoder contains seven blocks, each containing temporal convolutions used for dimensionality reduction and feature representation. There is a CNN layer responsible for modeling relative positional embeddings of the input sequence. As shown in Figure \ref{fig:wav2vec2}, the feature encoder, denoted as $f : X \rightarrow Z$, takes raw audio $X$ as input and creates latent speech representations (embeddings) $\boldsymbol{z}_{1}, \ldots,\boldsymbol{z}_{T}$ for $T$ time-steps. In the self-supervised pre-training phase, the output audio embeddings $\boldsymbol{z}_{1}, \ldots,\boldsymbol{z}_{T}$ from the feature encoder are discretized into a finite set of speech representations via a quantization module $Z \rightarrow Q$, resulting in discrete representations $\boldsymbol{q}_{t}$ for each time-step. The quantization step is removed when fine-tuning. 

These speech embeddings are then fed into the Transformers block, denoted as $g: Z \rightarrow C$, which produces contextualized representations $\boldsymbol{c}_{1}, \ldots,\boldsymbol{c}_{T}$ that capture information from the entire sequence.

\subsection{Pre-training Objective}

Pre-training wav2vec 2.0 consists of the joint optimization of a contrastive task $\mathcal{L}_{m}$ and a diversity task $\mathcal{L}_{d}$. Then, the objective $\mathcal{L}$ for the whole pre-training stage results in:

\begin{equation}
    \label{eq:wav2vec2_loss}
    \mathcal{L} = \mathcal{L}_{m} + \alpha\mathcal{L}_{d},
\end{equation}
where $\alpha$ is a tuned hyperparameter. 

The contrastive task $\mathcal{L}_{m}$ measures the discrepancy between the context vectors $\boldsymbol{c}_{t}$ and the true quantized representation $\boldsymbol{q}_{t}$ selected from a set of $K + 1$ quantized candidate representations, where $K$ representations are distractors. In contrast, the diversity task $\mathcal{L}_{d}$ encourages the model to use the codebook entries equally often, increasing the use of quantized codebook representations. Further details about wav2vec 2.0 pre-training objective can be found in \cite{dieleman2018challenge, baevski2020wav2vec}.

In this project, we explore using wav2vec 2.0 pre-trained with mono-channel speech signals, but we also adapt the framework to pre-train on multichannel spatial audio.  

\subsection{Fine-tuning wav2vec 2.0}

Fine-tuning the pre-trained wav2Vec 2.0 model for different speech-related tasks involves adapting the model's representations and hyper-parameters to the specific requirements of the downstream applications. A task-specific head or a linear layer on top of the base model is typically added during fine-tuning. The model learns these new specific weights by training on a smaller and labeled dataset specific to the task. 

The fine-tuning loss hardly depends on the task. For example, for Automatic Speech Recognition (ASR), the model may use the Connectionist Temporal Classification (CTC) approach \cite{graves2006connectionist} for audio-text alignment, while for speaker identification, fine-tuning might involve contrastive loss functions, where the model learns to discriminate between different speakers. Our proposal is to fine-tune wav2vec 2.0 for both the SED and DOA estimation tasks. We elaborate on the details of the fine-tuning process of our task in Section \ref{sec:method}.

\section{Methodology}\label{sec:method}

In this Section, we outline the methodology employed in our project to develop a self-supervised approach for SELD utilizing the latest version of the wav2vec 2.0 pre-training framework. Our model, named w2v-SELD, adapts both pre-training and fine-tuning frameworks of wav2vec 2.0 to work with spatial audio that contains four audio channels. Pre-training on large amounts of unlabeled 3D audio data takes us a step closer to overcoming the limitations posed by the traditional supervised learning methods, which require large amounts of annotated data to train. By leveraging this pre-training approach, our project aims to develop a highly accurate and robust SELD model fine-tuned on labeled spatial audio datasets with a restricted amount of recordings. In this Section, we describe the pre-training and fine-tuning stages of w2v-SELD and the data augmentation techniques applied for spatial audio.

\subsection{Model adaptation}

The feature encoder of our w2v-SELD model takes four channels of raw audio data as input and outputs feature vectors. The audio recordings follow the Ambisonics B-format for spatial audio \cite{zotter2019}, are re-sampled into \unit[16]{KHz}, and standardized to have zero mean and unit variance. We modified the first of the seven convolutional blocks of the feature encoder to accommodate a multichannel input. The remaining blocks follow the architecture of wav2vec 2.0. The output contains a series of feature vectors, where each one represents a segment of \unit[20]{ms} of input audio. The same feature encoder architecture of w2v-SELD was used for pre-training and fine-tuning. 

The Transformer blocks, following the feature encoder, incorporate an encoder Transformer architecture \cite{vaswani2017attention} with multi-head self-attention mechanisms. Each attention head is a separate mechanism that learns to attend to different parts of the input data independently, capturing various relationships and dependencies in the data. As in the original wav2vec 2.0, we experiment with two different setups: the BASE model, which contains 12 transformer blocks with 8 attention heads that output an embedding of dimension 768, and a LARGE model with 24 transformer blocks each one with 16 attention heads that output an embedding of dimension 1,024. 



\subsection{Pre-training}

We compare the performance of the w2v-SELD model for the SELD task by using the weights pre-trained on spatial or non-spatial audio. As for the non-spatial audio pre-training, the BASE wav2vec 2.0 model was pre-trained with the LibriSpeech (LS-960) \cite{panayotov2015librispeech} dataset which contains \unit[960]{hours}, while the LARGE  wav2vec 2.0 model was pre-trained with LibriVox (LV-60k) \cite{panayotov2015librispeech}, comprising a vast \unit[60,000]{hours} of audio. These datasets encompass English speech recordings featuring American, Canadian, and British accents. For the pre-training on spatial audio, we employed the TAU Spatial Sound Events datasets and the Learning 3D Audio Sources (L3DAS) datasets, which will be further detailed in Section \ref{datasets}. It is noteworthy that, as unsupervised pre-training does not require annotations, we could use spatial audio datasets annotated at different temporal resolutions.  

During pre-training, the model acquired an understanding of audio data structure by predicting masked frames within the input. This masking process involves the random selection and concealment of a subset of audio frames. In essence, certain segments of the input audio remain hidden from the model. To determine which time-steps to mask within the input sequence, each latent speech representation within an utterance is considered as a potential starting time-step with a probability of $p$, where $M$ denotes the length of each masked span from the respective time step. We adhere to the $p$ and $M$ values as described in \cite{baevski2020wav2vec} for pre-training w2v-SELD. 

By using mono-channel audio, the model is expected to capture various aspects of audio structure such as phonemes, prosody, and background noise, without the need for explicit phonetic transcriptions or annotations. Conversely, when employing spatial audio for pre-training, the model is expected it to learn additional spatial information from the input's 4 channels.

\subsection{Fine-tuning}

The different pre-trained versions of w2v-SELD are fine-tuned by incorporating two randomly initialized dense layers within the Transformer output---one dedicated to each sub-task of SELD. The SED branch is responsible for  multi-class classification, while the DOA branch performs a regression task. Adhering to the minimum temporal resolution of \unit[100]{ms} established for the DCASE Challenges focused on SELD, our model is designed to estimate SED classifications and DOA estimations at intervals of at least \unit[100]{ms}. 

For the fine-tuning process of w2v-SELD predictions, we explore two distinct approaches: frame-based prediction and sequence-based prediction. These methodologies will be further elaborated below. 

\subsubsection{Segment-based Prediction (w2v-SELD-SegPred)}

\begin{figure}[!htb]
    \centering    
    \includegraphics[width=0.5\linewidth]{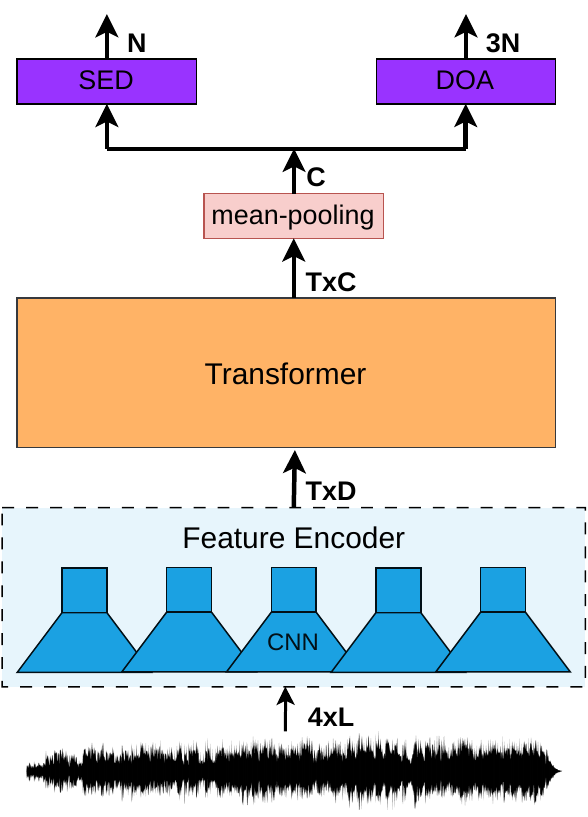}
    \caption{\label{fig:w2v_sed_doa_seqclass} Illustration of the w2v-SELD-SegPred approach. $N$ denotes the number of SED classes, $T$ represents the number of time-steps, and $C$ denotes the  embedding dimension of the w2v-SELD model.}
\end{figure}

In this approach, each spatial audio recording is segmented into \unit[100]{ms} intervals. The w2v-SELD model is then fed four channels of \unit[100]{ms} each and outputs a single SED and DOA prediction, as illustrated in Figure \ref{fig:w2v_sed_doa_seqclass}. The output from the Transformer blocks forms a 2D vector, which passes by a mean-pooling layer applied to the temporal dimension. Subsequently, the resulting 1D vector feeds two separate dense layers, each one for SED and DOA tasks. The dimension of the SED vector, denoted as $N$, corresponds to the number of sound classes present in the dataset. On the other hand, the DOA vector's dimension is $3N$, representing the Cartesian coordinates associated with each sound class. 

\subsubsection{Frame-based Prediction (w2v-SELD-FramePred)}

The second approach, named w2v-SELD-FramePred, receives a segment of 3D audio of \unit[2.97]{s} and generates predictions on a per-frame basis. Unlike the w2v-SELD-SegPred method, w2v-SELD-FramePred retains the original temporal dimension $T$ of the embedding. It predicts a vector for each time step $t=1,\ldots,T$. The choice of input audio duration and frame-wise prediction draws inspiration from SELDnet \cite{adavanne2018sound}.  However, our w2v-SELD-FramePred approach improves the time resolution of predictions to \unit[20]{ms}, which results from the CNN blocks within the feature encoder of the w2v-SELD model. This increased temporal resolution serves as an advantage as the model predicts outputs within smaller time windows, adhering to the minimum temporal resolution specified in the DCASE challenges. 

\begin{figure}[!htb]
    \centering    
    \includegraphics[width=0.5\linewidth]{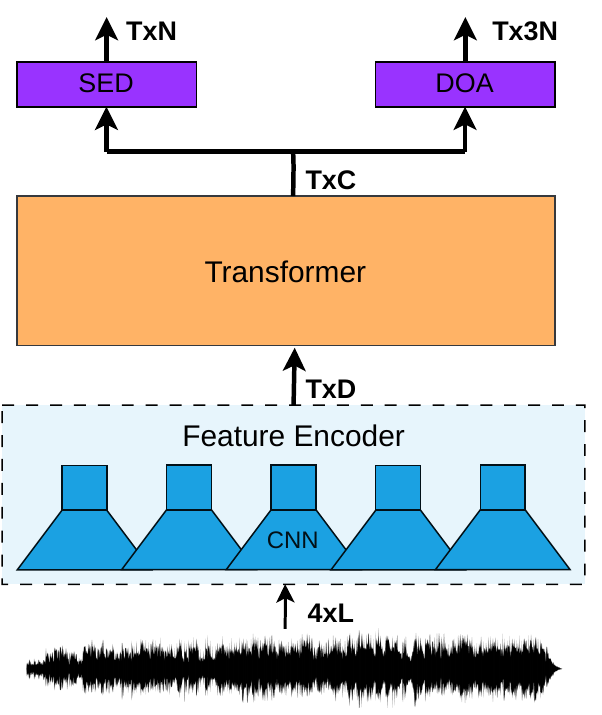}
    \caption{\label{fig:w2v_sed_doa} Illustration of the w2v-SELD-FramePred. $N$ denotes the number of SED classes, $T$ represents the number of time-steps, and $C$ denotes the embedding dimension of the w2v-SELD model.}
\end{figure}

As depicted in Figure \ref{fig:w2v_sed_doa}, the 2D vector embedding of the w2v-SELD model is fed into both the SED and DOA branches. Subsequently, the SED branch produces a 2D vector with dimensions $T \times N$, while the DOA branch yields a 2D vector of dimension $T \times 3N$. This approach adeptly captures the temporal correlations among sound events within the input segment.

\subsection{Data Augmentation for Spatial Audio}\label{DA_seld}

Data augmentation techniques for audio signals are used to create variations in audio data, enhancing the robustness and generalization of deep learning models. Beyond conventional data augmentation methods, spatial audio benefits from specific techniques tailored for multichannel inputs and the spatialization encoded in the Ambisonics B-format. Augmenting our training samples plays a crucial role, especially during the fine-tuning phase when dealing with limited datasets of 3D audio. The data augmentation techniques applied during the fine-tuning stage of this study are described below.

\subsubsection{Traditional techniques}

We apply traditional techniques individually to each channel of spatial audio. Leveraging the Torch-Audiomentations library \cite{pariente2020asteroid}, we use methods such as random gain, colored and background noise addition, and pitch shift. These modifications are directly applied to the audio waveform, while the SED and DOA annotations are unmodified. 

\subsubsection{Time and frequency masking}

\begin{figure}

\begin{minipage}{.5\linewidth}
\centering
\subfloat[]{\label{main:a}\includegraphics[width=\linewidth]{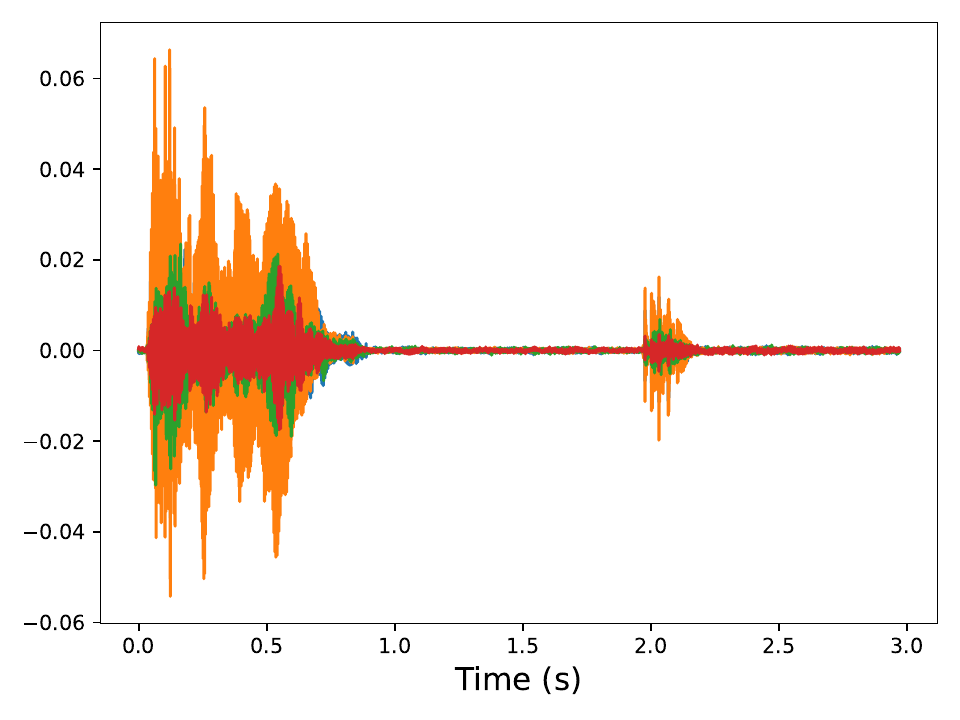}}
\end{minipage}%
\begin{minipage}{.5\linewidth}
\centering
\subfloat[]{\label{main:b}\includegraphics[width=\linewidth]{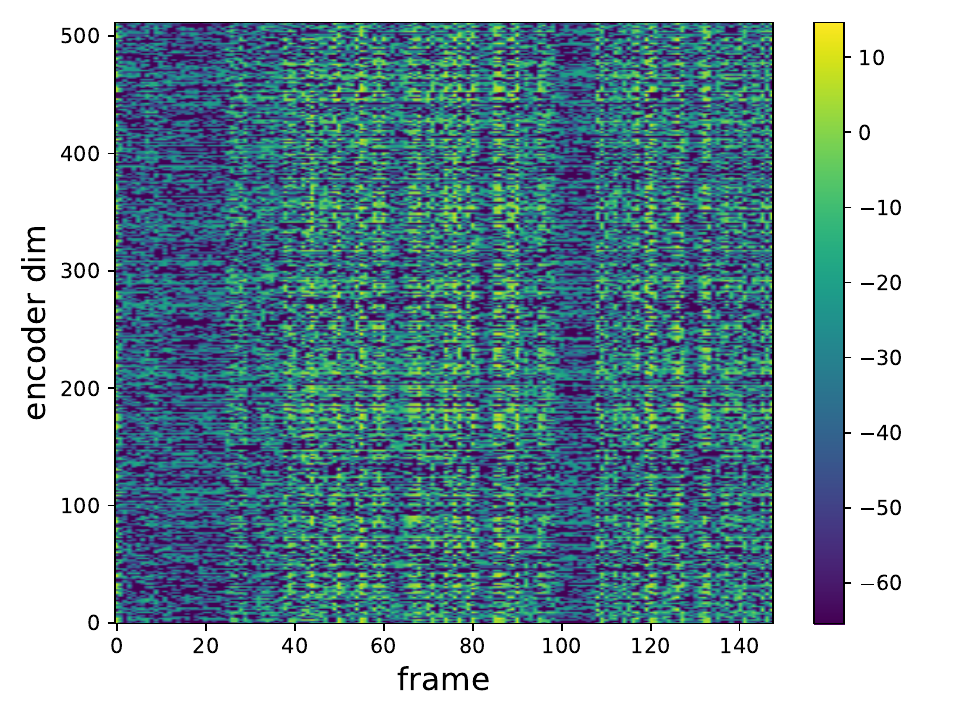}}
\label{spec2}
\end{minipage}
\centering
\subfloat[]{\label{main:c}\includegraphics[width=0.6\linewidth]{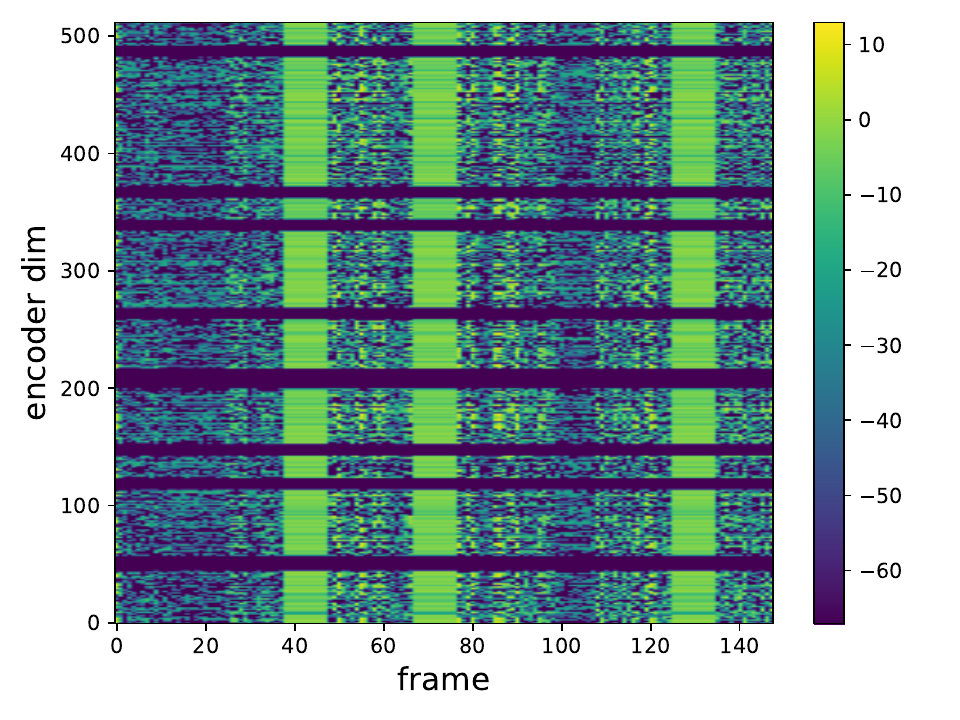}}
\caption{SpecAugment on w2v-SELD encoder output. (a) Raw audio signal. (b) Feature encoder output. (c) Feature encoder output with SpecAugment.}
\label{fig:spec_augment}
\end{figure}

Time and frequency masking techniques, as proposed in the SpecAugment library \cite{park2019specaugment}, are employed. SpecAugment originally operates on a spectrogram, masking random time frames and frequency bins. However, in our case, we implement this technique on the output of the feature encoder, which encompasses time and frequency dimensions. The frequency dimension is replaced by the 512 channels outputted by the CNN blocks. This process is illustrated in Figure \ref{fig:spec_augment}, where a multichannel audio waveform is processed by the feature encoder to obtain an embedding representation (Figure \ref{fig:spec_augment}b) where random time and frequency masking is applied. Figure \ref{fig:spec_augment}c shows the resulting masked embedding, which is then fed into the Transformer blocks.

\subsubsection{Sound Field Rotation}

The B-format is a standardized representation of Ambisonics audio that encodes information about the acoustic scene in all directions. The multichannel representation comprises one omnidirectional channel (W) and three channels-- X, Y, and Z--encoding pressure differences between hemispheres, expressed in Cartesian coordinates. 

Channel swapping is a technique that rotates the sound field in spatial audio recordings. This technique alters the perceived localization of sounds by swapping audio channels and reversing their signs. These modifications are used to transform the DOA annotations, ensuring congruence with the sound field rotation. Eight types of transformations, as outlined in \cite{huang2021sseldnet}, are randomly applied during training in the fine-tuning stage.

\section{Experimental Setup}\label{sec:setup}



The w2v-SELD model was pre-trained on an A100 GPU for $3$ days for the BASE version and $7$ days for the LARGE version. For fine-tuning, any GPU with at least \unit[16]{GB} of RAM can be employed. In this project, we fine-tuned the w2v-SELD model using an NVIDIA TITAN V GPU, which typically requires approximately \unit[10]{hours} to complete the process.   

The growing interest in 3D audio analysis has been accompanied by an increase in the number of spatial audio datasets. This expansion has been facilitated by advancements in recording and playback devices able to handle spatial audio. However, the limited availability of datasets is a result of the challenge of collecting accurate annotations for SED and DOA. In this study, we combine multiple 3D audio datasets for pre-training the model, while two smaller datasets were used for fine-tuning. The following section provides a brief overview of the characteristics of the spatial audio datasets considered for this study.

\subsection{Spatial Audio Datasets}\label{datasets}

The spatial audio datasets used in the pre-training of the w2v-SELD model are described below. 

\subsubsection{L3DAS datasets}

The L3DAS21 \cite{guizzo2021l3das21} and L3DAS22 \cite{guizzo2022l3das22} challenges focus on 3D speech enhancement (SE) and 3D SELD. These datasets were created by convolving mono-channel sound events from FSD50K~\cite{fsd50k} or speech from LibriSpeech with impulse responses (IRs) captured using two First-Order Ambisonic (FOA) microphones. Each dataset contains two subsets: one designed for SE and the other for SELD. In the SE subset, each recording contains speech, whereas, in the SELD subset, voiced sounds including speech may be present. While only the subsets for SELD provide annotations for SED and DOA, the SE subsets can be used along with the SELD subset during the unsupervised pre-training stage of our model, as annotations are unnecessary for this stage; only Ambisonic recordings are required. 

The L3DAS21 dataset comprises approximately \unit[65]{hours} of Ambisonics audio recordings, distributed into \unit[50]{hours} dedicated to SE and \unit[15]{hours} for SELD. To synthesize the 3D audio, 14 types of sounds typically encountered in an office environment were convolved with the IRs. Clean sound samples were sourced from the LibriSpeech and FSD50K~\cite{fsd50k} datasets, while four types of office-like background noises were chosen from FSD50K. The sampling frequencies for the SE and SELD subsets are \unit[16]{kHz} and \unit[32]{kHz}, respectively.

In contrast, L3DAS22, an extension of the L3DAS21 dataset, offers an expanded dataset with an increased volume of data. It comprises approximately \unit[94]{hours} of audio recordings, with \unit[86]{hours} allocated for the SE task and \unit[7.5]{hours} for SELD. The type of sounds, background noises, and recording strategies remain consistent with those of the L3DAS21 dataset.

\subsubsection{Tampere University datasets}

The Tampere University of Technology (TUT) released a series of SELD datasets specifically designed for use in the DCASE competitions. All these datasets contain recordings in the FOA B-format for spatial audio, alongside annotations for SED and DOA. In this study, we consider the datasets released annually from 2018 to 2021.  

TUT-2018 \cite{adavanne2018sound}: This dataset incorporates stationary sound events placed synthetically in specific spatial coordinates. For this study, we employ the ANSYN and REAL datasets from TUT-2018, which were synthesized using artificial and real IRs, respectively. The ANSYN dataset simulates an anechoic environment without reverberation, while the REAL dataset uses IRs recorded using a spherical microphone array in a university corridor surrounded by classrooms. Sound events isolated from the urbansound8k dataset \cite{UrbanSound} were used in both datasets. REAL comprises 10 types of sounds, while ANSYN considers 11 classes. Each dataset contains up to three overlapping sound events in the recordings. There are 2,700 spatial audio files in each dataset, sampled at \unit[44.1]{kHz} for \unit[30]{s}, totaling \unit[22.5]{hours} in each dataset.

TAU-2019 \cite{Adavanne2019}: This dataset was synthesized by convolving 11 types of static sound events with real-life IRs collected from five rooms with different reverberant characteristics at multiple spatial coordinates. The sound events remain stationary with fixed locations throughout their duration. The dataset comprises two sub-datasets: TAU Spatial Sound Events 2019 - Ambisonic (FOA), and TAU Spatial Sound Events 2019 - Microphone Array (MIC).  The FOA follows the B-format of Ambisonics, while MIC uses a tetrahedral microphone array to capture the IRs. Each sub-dataset's development set includes 400 one-minute-long recordings sampled at \unit[48]{kHz}, and the evaluation set consists of 100 one-minute recordings. Approximately \unit[8]{hours} of spatial audio are available in this dataset.

TAU-2020 \cite{Politis2020_task3_report}: This dataset introduced diversified acoustical conditions and sound event trajectories. It contains 714 sound examples of 14 categories convolved with recorded IRs, encompassing static and dynamic trajectories. Similar to TAU-2019, FOA and MIC sub-datasets were released. Each sound event in the sound scene is linked to its DOA trajectory, onset, and offset times. The development set includes 600 one-minute-long recordings sampled at \unit[24]{kHz}, and the evaluation set comprises 200 one-minute recordings. The dataset offers around \unit[13]{hours} of spatial audio.

TAU-2021 \cite{Politis2020_task3_report}: Released for the DCASE2021 Challenge, this dataset features moving sound sources, directional interference events, and an additional layer of background noise in all samples. It includes 12 classes of sound events and realistic spatialization and reverberation achieved through IRs collected in 13 different enclosures. The development dataset consists of 500 one-minute-long audio samples at a sampling rate of \unit[24]{kHz}, while the testing set contains 100 one-minute-long audio samples. This dataset comprises \unit[10]{hours} of spatial audio. 

\subsection{Pre-training hyper-parameters}

Adhering to the pre-training methodology of wav2vec 2.0, wherein the LARGE model employs a more extensive dataset compared to the BASE model, we integrate the L3DAS22 dataset to augment the dataset size for training our LARGE w2v-SELD model. Generally, we combine unlabeled 3D audio data for pre-training the w2v-SELD model as follows:

\begin{itemize}
    \item \textbf{L3DAS21-SELD}: Combination of L3DAS21, TUT-2018, TAU-2019, TAU-2020, and TAU-2021 datasets.
    \item \textbf{L3DAS22-SELD}: Combination of L3DAS22, TUT-2018, TAU-2019, TAU-2020, and TAU-2021 datasets.
\end{itemize}

We downsample the 3D audio signals to a \unit[16]{kHz}
frequency and subsequently divide them into frames lasting \unit[4]{s}. The BASE w2v-SELD model was trained using the L3DAS21-SELD set, resulting in $179,442$ utterances, while the LARGE w2v-SELD model utilized the L3DAS22-SELD set, comprising $227,797$ utterances after preprocessing.

The original wav2vec 2.0 pre-training stage uses audio segments lasting \unit[15.6]{s}. 
 However, we have reduced the duration of our segments. This adjustment was made due to the heightened complexity of learning unsupervised representations for SED and DOA from multichannel audio, which presents a more challenging task compared to the ASR goal of wav2vec 2.0.



In the w2v-SELD pre-training, the model is optimized by initially warming up the learning rate for the first $32,000$ updates, reaching a peak of $5 \times 10^{-4}$ for both BASE and LARGE models, followed by a linear decay. The BASE model was pre-trained using the L3DAS21-SELD dataset for $400\text{k}$ updates, while the LARGE model was pre-trained on the L3DAS22-SELD dataset for $600\text{k}$ updates.

\subsection{Fine-tuning hyper-parameters}

The layer-wise learning rate employs three distinct rates for different components: the w2v-SELD model until the Transformer's output, SED, and DOA branches. Additionally, a tri-stage learning rate schedule is implemented to cautiously adapt to the new dataset without compromising the knowledge gained during pre-training. To summarize, the following hyper-parameters are configured: 



\begin{itemize}
    \item The tri-stage schedule comprises  10\% warm-up, 30\% constant learning, and 60\% decay phases. 
    \item The learning rates are set as follows: $5 \times 10^{-5}$ for the w2v-SELD model until the Transformer's output, $5 \times 10^{-4}$ for SED and DOA branches.
\end{itemize}

As observed, the learning rate for the w2v-SELD model until the Transformer's output is lower than that of the SED and DOA branches. This intentional discrepancy aims to leverage the knowledge acquired during the pre-training stage.

\subsection{Evaluation Metrics}

The evaluation metrics used for SELD gauge the performance of systems based on their ability to accurately detect and localize sound events. We have adopted segmented-based metrics as outlined in \cite{mesaros2016metrics}, commonly referred to as frame-based metrics. For SED, the metrics used are F-score (F1) and Error Rate (ER). Regarding the DOA metrics, we use the DOA\textsubscript{error} to measure the difference between the estimated and reference DOA angle, and the Frame-Recall (FR) defined as the relation between true positive and true negative estimations.
The overall score, named SELD\textsubscript{score}, combines both SED and DOA metrics as follows:
\begin{equation}
    \text{SELD\textsubscript{score}} = \frac{\text{SED\textsubscript{score}}+\text{DOA\textsubscript{score}}}{2},
\end{equation}
\noindent where
\begin{gather} 
\text{SED\textsubscript{score}} = \frac{\text{ER}+ (1 -\text{F1})}{2}, \\
\text{DOA\textsubscript{score}} = \frac{\text{DOA\textsubscript{score}/180}+ (1 -\text{FR})}{2}.
\end{gather}

In an ideal scenario, the metrics for a perfect SELD model would be $\text{F1}=100\%$, $\text{ER}=0$, $\text{DOA-error}=0$, and $\text{FR}=100\%$, resulting in a $\text{SED-score}$ of $0$. 

In the DCASE2020 Challenge, the SED metrics (i.e., F1 and ER) consider a frame prediction as a true positive if the spatial error, calculated as the angular distance between reference and predicted DOA, is less than $20^{\circ}$ \cite{Politis2020_task3_report}.

\section{Results}\label{sec:results}

In the following sections, we present the results of our proposed w2v-SELD approach, fine-tuned using the TAU-2019 and TAU-2020 datasets. Initially, we examine the influence of employing the w2v-SELD-FramePred approach in contrast to the w2v-SELD-SegPred approach (Section \ref{seq_vs_frame}). Subsequently, in Section \ref{pret_exps}, we assess the effects of pre-training our w2v-SELD model using various unlabeled datasets. This section offers a comparison between model performance during linear evaluation with frozen parameters and when parameters are updated during fine-tuning. Moving forward, Section \ref{sota_comp} compares our approaches against baseline systems and state-of-the-art techniques. Finally, we evaluate the impact of data augmentation on fine-tuning the SELD task (Section \ref{DA_impact}).


\subsection{Sequence Classification versus Audio Frame Classification} \label{seq_vs_frame}

The primary objective of this experiment is to compare the performance between the w2v-SELD-FramePred and w2v-SELD-SegPred approaches, regardless of the type of audio used for pre-training. To accomplish this, we implemented both strategies on two models: the wav2vec 2.0 BASE model, pre-trained on mono-channel audio data, and our w2v-SELD model, pre-trained on spatial audio data. Subsequently, all models underwent fine-tuning using the TAU-2019 dataset.

\begin{table}[!ht]
		\centering
		\caption{Comparison between prediction strategies: w2v-SELD-FramePred and w2v-SELD-SegPred fine-tuned using the TAU-2019 dataset.$^*$} 
		\resizebox{0.8\linewidth}{!}{%
			\begin{tabular}{cccccccc}
			    \hline  \hline
				  & \textbf{Strategy} & \textbf{Pre-training Set} &  $\downarrow$\textbf{ER} & $\uparrow$\textbf{F1}\% & $\downarrow$\textbf{DOA} & $\uparrow$\textbf{FR}\% & $\downarrow$\textbf{SELD}\\ 
				\hline
                    & w2v-SELD-SegPred & LS-960 & 1.12 & 54.90 & 17.08 & 72.84 & 0.49\\
				& w2v-SELD-SegPred & L3DAS21-SELD & 0.94 & 58.83 & 17.06 & 74.04 & 0.43
                    \\
                    & w2v-SELD-FramePred & LS-960 
                    & 0.14 & 91.5& \textbf{4.8} & 91.66 & 0.08 \\
                    & w2v-SELD-FramePred & L3DAS21-SELD & \textbf{0.10} & \textbf{94.20} & 4.88 & \textbf{93.12} & \textbf{0.06} \\
                    \hline
                    \hline
		\end{tabular}}
		\label{tab:results1}\\
  \tiny{$^*$The arrows indicate whether the metric improves with an increase ($\uparrow$) or decrease ($\downarrow$) in its value.}
\end{table}

The results presented in Table \ref{tab:results1} demonstrate the advantages of employing the w2v-SELD-FramePred approach across both pre-training sets.  The adoption of w2v-SELD-FramePred technique led to a significant enhancement in the SELD\textsubscript{score} by $82\%$ and $86\%$ for the LS-960 and L3DAS21-SELD pre-training sets, respectively. This remarkable improvement emphasizes the significance of leveraging temporal dependencies within the input representation. Despite using short frames lasting \unit[2.97]{s} for fine-tuning, the w2v-SELD-FramePred approach effectively utilizes previous information to enhance future predictions. Conversely, the wav2vec2-SeqPred approach loses temporal relationships among samples, resulting in a detrimental impact on the SELD system's performance. Although wav2vec2-SeqPred might seem the more intuitive choice for the SELD task, it consistently fails to yield optimal results, regardless of the pre-training data employed.

\subsection{Impact of Pre-training on Spatial Audio} \label{pret_exps}

Upon evaluating our prior analyses, the audio frame prediction approach (i.e., w2v-SELD-FramePred) emerged as the most effective method for both sound event classification and localization. Consequently, this approach is employed in the subsequent experiments. To further assess the influence of pre-training on the SELD task, a series of fine-tuning experiments were conducted on the TAU-2019 dataset using weights acquired from various pre-training configurations of our w2v-SELD model. During fine-tuning, the parameters of w2v-SELD are not back-propagated, similar to a model in inference mode. With our focus on evaluating the representations learned during pre-training using a linear evaluation, all layers of the w2v-SELD model before the SED and DOA branches are frozen. This ensures that only the SED and DOA layers' weights are updated during fine-tuning, while the preceding layers remain isolated from this process.

We experiment with three configurations: 1) no pre-training, for which the w2v-SELD model's weights are randomly initialized and frozen during fine-tuning, 2) pre-training on LS-960 mono-channel speech data, and 3) pre-training on the L3DAS21-SELD set of unlabeled spatial audio data. It is essential to note that across these  configurations, the weights of the w2v-SELD model before the SED and DOA branches remain frozen, and only the SED and DOA layers are updated during fine-tuning. These experiments aim to gain a deeper understanding of the impact of the pre-training stage on the performance of the SELD task and to identify any potential areas for improvement.


Table \ref{tab:results2} exhibits the results for the mentioned configurations. These metrics are not expected to reach optimal values, as our objective here is to evaluate the pre-training representations rather than optimal performance. Notably, we observed a consistent trend of improvement with pre-training, even when utilizing mono-channel data, compared to fine-tuning the SED and DOA branches from scratch. Pre-training the model on the LS-960 unlabeled speech dataset improved the SELD\textsubscript{score} by 10\%. The use of spatial audio for pre-training led to a 23\% SELD\textsubscript{score} improvement compared to mono-channel data. Moreover, significant improvements in SED metrics were observed, with a 55.28\% increase in F1 when pre-training on L3DAS21-SELD data compared to the LS-960 dataset, and a 144.43\% improvement compared to the model without pre-training. Additionally, it is evident that the DOA metrics are influenced by updating the parameters of the w2v-SELD model during fine-tuning, which significantly contributes to their improvement.

\begin{table}[!ht]
		\centering
            \caption{Linear evaluation of the frozen parameters of the w2v-SELD model pre-trained various datasets. Evaluation conducted on the TAU-2019 dataset.} 
		\resizebox{0.8\linewidth}{!}{%
			\begin{tabular}{cccccccc}
			    \hline  \hline
                    & \textbf{Pre-training Set} & $\downarrow$\textbf{ER} & 	$\uparrow$\textbf{F1}\% & $\downarrow$\textbf{DOA-error} & $\uparrow$\textbf{FR}\%  & $\downarrow$\textbf{SELD\textsubscript{score}} \\ 
				\hline
                    & - & 0.90 & 26.67 & \textbf{70.74} & 50.27 & 0.63\\
                    &  { \footnotesize LS-960 } & 0.81 & 41.98 & 72.15 & 52.96  & 0.57\\
                    & { \footnotesize L3DAS21-SELD } & \textbf{0.56} & \textbf{65.19} & 80.06 & \textbf{58.45} & \textbf{0.44}\\
                    \hline
                    \hline
		\end{tabular}}
		\label{tab:results2}\\
  \tiny{$^*$The arrows indicate whether the metric improves with an increase ($\uparrow$) or decrease ($\downarrow$) in its value.}
\end{table}



Subsequently, we fine-tuned  the same pre-training configurations but unfreezing the w2v-SELD model parameters. This allows the initialized pre-trained weights to be updated during fine-tuning for improved adaptation to the SELD task. Table \ref{tab:results3} highlights the advantageous outcomes of the proposed w2v-SELD-FramePred approach, showcasing a $20\%$ SELD\textsubscript{score} enhancement when pre-trained on the LS-960 dataset and a $40\%$ improvement using the L3DAS21-SELD spatial audio set for pre-training, compared to no pre-trained weights.


\begin{table}[!ht]
		\centering
		\caption{Performance results of the w2v-SELD BASE model pre-trained using various datasets and fine-tuned on the TAU-2019 dataset.}
		\resizebox{0.8\linewidth}{!}{%
			\begin{tabular}{ccccccc}
			    \hline  \hline
                    & \textbf{Pre-training Set} & $\downarrow$\textbf{ER} & 	$\uparrow$\textbf{F1}\% & $\downarrow$\textbf{DOA-error} & $\uparrow$\textbf{FR}\%  & $\downarrow$\textbf{SELD\textsubscript{score}} \\ 
				\hline
                    &  - &  0.17 & 89.45 & 4.97 & 92.01 & 0.10\\
                    &  { \footnotesize LS-960}  &  0.14 & 91.50 & \textbf{4.80} & 91.66 & 0.08\\
                    & { \footnotesize L3DAS21-SELD} & \textbf{0.10} & \textbf{94.20} & 4.88 & \textbf{93.12} & \textbf{0.06}
                    \\
                    \hline
                    \hline
		\end{tabular}}
		\label{tab:results3}\\
  \tiny{$^*$The arrows indicate whether the metric improves with an increase ($\uparrow$) or decrease ($\downarrow$) in its value.}
\end{table}

\subsection{Comparison with state-of-the-art models}\label{sota_comp}

The proposed w2v-SELD model, implemented with the w2v-SELD-FramePred approach, was fine-tuned and subsequently evaluated on both the TAU-2019 and TAU-2020 datasets. Our approach was compared against the baseline models provided for each dataset, as well as the current state-of-the-art models. We present the summarized results for both our BASE and LARGE w2v-SELD models, pre-trained on the L3DAS21-SELD and L3DAS22-SELD datasets, respectively, in Table \ref{tab:results4}.

\renewcommand{\arraystretch}{1.2}
\begin{table}[!ht]
		\centering
		\caption{Comparison of w2v-SELD-FramePred performance against baseline and state-of-the-art methods.}
		\resizebox{\linewidth}{!}{%
			\begin{tabular}{cccccccc}
			    \hline  \hline
				  Dataset & model & Unlabeled Data & $\downarrow$\textbf{ER} & $\uparrow$\textbf{F1} \% & $\downarrow$\textbf{DOA-error} & $\uparrow$\textbf{FR} \% & $\downarrow$\textbf{SELD\textsubscript{score}}\\ 
				\hline
                    \multirow{8}{*}{TAU-2019}
                    & { \footnotesize CRNN (baseline)} \cite{Adavanne2019} & - & 0.28 & 85.40& 24.60 &  85.40 & 0.18\\ 
                    & { \footnotesize Ensemble-CRNN} \cite{kapka2019sound}  & - & \textit{0.08} &  \textit{94.70} & \textit{3.70} & \textit{96.80} & \textit{0.06}\\ 
                    \cdashline{2-8}
                    & \textbf{This work} \\
                    & { \footnotesize BASE} &  LS-960 &  0.142 & 91.50 & 4.80 & 91.66 & 0.080\\
				& { \footnotesize LARGE} 
                    & LV-60k & 0.139 & 92.04 & 4.84 & 92.77 & 0.079 \\
                    & { \footnotesize BASE} & L3DAS21-SELD & 0.099 & 94.20 & 4.88 & \textbf{93.12} & 0.063
                    \\
                    & { \footnotesize LARGE} & L3DAS22-SELD & \textbf{0.096} & \textbf{94.66} & \textbf{4.67} & 93.05  & \textbf{0.061}
                    \\
                    \hline
                    \multirow{8}{*}{TAU-2020
                    }
                    & { \footnotesize CRNN (baseline)} \cite{Politis2020_task3_report} &  - & 0.69 &  41.30 & 23.10 & 62.40 & 0.45\\ 
                     & { \footnotesize Ensemble-CRNN} \cite{Du2020_task3_report} & - &
                     \textit{0.20} & \textit{84.90} & \textit{6.00} & \textit{88.50} & \textit{0.13}\\ 
                    \cdashline{2-8}
                    & \textbf{This work} \\
                    & { \footnotesize BASE} &  LS-960 & 0.363 & 72.56 & 9.06 & 76.43 & 0.231\\  
				& { \footnotesize LARGE} & LV-60k &  0.378 & 70.77 & 10.68 & 76.43 & 0.241
                    \\
                    & { \footnotesize BASE} & L3DAS21-SELD & 0.307 & 76.96 & \textbf{8.37} & 79.96 & 0.196\\
                    & { \footnotesize LARGE} & L3DAS22-SELD  & \textbf{0.301} & \textbf{77.17} & 8.51 & \textbf{80.83} & \textbf{0.192}\\
                    \hline
                    \hline
		\end{tabular}}
		\label{tab:results4}\\
  \tiny{$^*$The arrows indicate whether the metric improves with an increase ($\uparrow$) or decrease ($\downarrow$) in its value. Additionally, we have highlighted in bold the metrics of the best approach observed among the experiments conducted in this study for each dataset. In italic is highlighted the state-of-art metric obtained in each dataset.}
\end{table}

On TAU-2019, all models trained in this study outperformed the baseline system \cite{Adavanne2019}. Notably, our best model (LARGE/L3DAS22-SELD) exhibited a 66\% improvement in SELD\textsubscript{score} compared to the baseline. Moreover, our top-performing model achieved an SELD\textsubscript{score} very close to the state-of-the-art system for TAU-2019 \cite{kapka2019sound}, trailing behind by only 1.64\%.

Moving to TAU-2020, the SED metrics were adjusted to align with the DCASE2020 Challenge criteria, where a spatial error of less than $20^{\circ}$ was required to classify as a true positive. Once again, all our experiments surpassed the baseline system for TAU-2020 \cite{Politis2020_task3_report} in both SED and DOA metrics. Our best model (LARGE/L3DAS22-SELD) exhibited a 57\% improvement in overall SELD\textsubscript{score} compared to the baseline system. However, there remains room for enhancement in comparison to the state-of-the-art system proposed in \cite{Du2020_task3_report}, where our best approach lags by $32\%$ in the SELD\textsubscript{score}.

We highlight the fact that the significant performance improvements in our approach primarily stemmed from the pre-training stage using unlabeled multichannel audio. In contrast, the state-of-the-art system proposed in \cite{Du2020_task3_report} heavily relies on manually synthesizing new Ambisonics audio samples by mixing non-overlapping samples and transforming them into spatial audio format. Given that synthesized samples require manual labeling, this process is time-consuming and challenging to replicate. Additionally, the system proposed in \cite{Du2020_task3_report} does not provide either a repository or the data used for training the model. 

We emphasize the availability of a repository of our proposed approach, where the trained models can be utilized for SELD task inference or the source code can serve as a foundational framework for further enhancements.

\subsection{Impact of data augmentation for spatial audio}\label{DA_impact}

As evinced in \cite{Du2020_task3_report}, the use of data augmentation holds significant importance in enhancing the SELD task, particularly in reducing  DOA metrics. In light of this, we conducted a further investigation to assess the impact of employing data augmentation techniques, detailed in Section \ref{DA_seld}, on the performance of our best approach (LARGE/L3DAS22-SELD). The results comparing the SELD metrics with and without data augmentation for the best models on both TAU-2019 and TAU-2020 are presented in Table \ref{tab:results5}. 

\begin{table}[H]
		\centering
		\caption{Comparison of top-performing w2v-SELD-FramePred model with and without data augmentation.} 
		\resizebox{0.8\linewidth}{!}{%
			\begin{tabular}{cccccccc}
			    \hline  \hline
				  & Dataset & DA &  $\downarrow$\textbf{ER} & $\uparrow$\textbf{F1} & $\downarrow$\textbf{DOA\textsubscript{error}} & $\uparrow$\textbf{FR} & $\downarrow$\textbf{SELD\textsubscript{score}} \\ 
				\hline
                    & \multirow{2}{*}{TAU-2019} & \xmark & 0.106 & 94.29\% & 5.57 & \textbf{94.55}\% & 0.062 \\
				& & $\checkmark$
                    & \textbf{0.096} & \textbf{94.66}\% & \textbf{4.67} & 93.05\% & \textbf{0.061} \\
                    & \multirow{2}{*}{TAU-2020} & \xmark & 0.368 & 72.29\% & 11.18 & 78.97\% & 0.229
                    \\
                    &  & $\checkmark$ & \textbf{0.301} & \textbf{77.17}\% & \textbf{8.51} & \textbf{80.83}\% & \textbf{0.192} \\
                    \hline
                    \hline
		\end{tabular}}
		\label{tab:results5}
\end{table}

The inclusion of data augmentation notably improved the overall SELD\textsubscript{score} for both datasets. Specifically, enhancements of $1.6\%$ and $16.16\%$ were observed for TAU-2019 and TAU-2020, respectively. This demonstrates the relevance of augmenting labeled spatial audio datasets for effective fine-tuning purposes. While we implemented well-established data augmentation techniques, we acknowledge the potential for further exploration of additional methods, especially considering the limited DOAs in the development set. Exploring novel augmentation techniques might yield greater improvements in DOA metrics. However, we highlight the fact that the improvement in metrics introduced by our top-performing approaches is not solely dependent on data augmentation techniques during fine-tuning. Instead, our focus lies on leveraging unlabeled spatial audio, a more feasible approach when manual annotations are unavailable.  


\section{Discussion}\label{sec:discussion}

While there remains potential for further improvements in the DOA metrics (DOA\textsubscript{error} and FR) for both datasets, it is worth highlighting that our study has achieved promising localization metrics solely by utilizing raw audio as input. Significantly, we achieved this without reliance on spectrograms, phase, or intensity vectors. To the best of the authors' knowledge, this constitutes the first instance of such results in the field. Furthermore, our models were pre-trained on relatively small datasets, approximately $65$ and $94$ hours for the BASE and LARGE models, respectively, in contrast to speech recognition where the original wav2vec 2.0 was pre-trained with datasets of $960$ and $60,000$ hours.

The achievements of our study underline the potential for successful localization without intricate audio pre-processing techniques, due to the self-supervised pre-training stage. These findings not only open new research avenues but also emphasize the significance of our contribution to the field.

The wav2vec 2.0 framework demonstrates its task-agnostic nature concerning audio and speech data, extending beyond exclusive use in the speech recognition domain. This suggests that the pre-task learns comprehensive representations of audio data, transcending specific task boundaries.

\section{Conclusions}\label{sec:conclusion}

This paper presented several noteworthy contributions in the domain of SELD, building upon our findings. Firstly, we introduced an SSL approach specifically tailored for SELD, utilizing the wav2vec 2.0 pre-training framework. This novel method showcased the potential for robust SELD models without extensive reliance on labeled spatial audio datasets. We adapted the fine-tuning process of our w2v-SELD model to incorporate SED and DOA estimation at the frame-level, refining the precision and accuracy of predictions. Lastly, we conduct a comprehensive evaluation of our w2v-SELD model pre-trained on diverse datasets, offering insights into its adaptability and performance across various pre-training configurations. We verified a $20\%$ improvement in the SELD\textsubscript{score} metric when using pre-trained weights instead of training the model from scratch. Furthermore, pre-training w2v-SELD on unlabeled spatial audio improved the SELD\textsubscript{score} metric by $40\%$ compared with pre-training on mono-channel audio. Our w2v-SELD approach outperformed the baseline systems provided for each dataset, reaching a $66\%$ improvement in SELD\textsubscript{score} and achieving close state-of-the-art performance using raw spatial audio as input instead of relying on spectrograms, phase, or intensity vectors. Finally, this research not only advances the field of SELD but also holds potential implications for broader applications in audio processing and machine learning.

In future research endeavors, we intend to explore the use of the Conformer architecture specifically tailored for the SELD task. 
This proposed direction seeks to handle frequency and phase representations as input, which can contain a more informative representation compared to raw audio data. The adaptation of the model to different input formats will increase its potential for real-world applications and provide valuable insights into its performance.

\bibliographystyle{unsrtnat}
\bibliography{references}  






\section*{Acknowledgments}

\begin{itemize}
    \item This work used resources of the "Centro Nacional de Processamento de Alto Desempenho em São Paulo (CENAPAD-SP)."

    \item Roberto de Alencar Lotufo is partially supported by CNPq (The Brazilian National Council for Scientific and Technological Development) under grant 313047/2022-7.
\end{itemize}

\end{document}